\documentclass[aip,jap, reprint, floatfix]{revtex4-1}

\usepackage{amsmath}
\usepackage{graphicx}
\usepackage{textcomp}
\usepackage{url}

\DeclareMathOperator{\sgn}{sgn}

\begin{document}

\title{Stable spatially discrete envelope-function model for graphene-like band structures}
\thanks{The author wish to thank Roger K. Lake for directing his attention to this problem, and to Avik Ghosh, Liang Fu and Fan Zhang for useful discussions.}

\date{\today}

\author{William R. Frensley}
\affiliation{Electrical Engineering, University of Texas at Dallas, Richardson, TX 75080}
\email{frensley@utdallas.edu}

\begin{abstract}
  The spurious states found in numerical implementations of envelope function models for semiconductor heterostructures and 
nanostructures have been shown to be readily removed by employing a first-order difference scheme.  This approach is applied
to the band structure of graphene.  The massless Dirac equation is identical to a simple two-band model with zero energy gap.
A first-order discretization of this equation produces strictly monotonic dispersion 
relations with the desired linear dependence on $k$ near the origin, thus removing the ``fermion doubling'' anomaly associated 
with formulation on a computational mesh.  The first-order formulation produces an ambiguity in the form of the Hamiltonian;
both forms produce identical results for physical observeables, and are related by both unitary transformations and time reversal.  
Other details needed to evaluate the properties of current-carrying
systems, including current density expressions, open boundary conditions and the treatment of heterojunctions, are also developed.

\end{abstract}

\maketitle

\section{Introduction}

A popular, but problematic, approach to calculation of the electronic states in semiconductor heterostructures is the 
envelope-function formulation \cite{Luttinger1955}.  The problems have been the persistence of spurious states in 
the spatially discrete formulations required to numerically treat structures of realistic complexity
\cite{Cartoixa2003,Vogl2011}.  These spurious states are simply an artifact of the use of the centered-difference
approximation to the gradient operator implied by linear $k$ terms.  The centered-difference leads to a non-monotonic
dispersion relation, which introduces spurious states with $k$ near the numerical Brillouin zone boundary, but with
energies that coincide with the states one wishes to evaluate.  

The solution to this problem is to simply abandon the use of the centered-difference and to instead formulate the 
linear-$k$ terms as one-sided, or nearest-neighbor, differences 
\cite{Bimberg1997,Jiang2014,Ma2014,Frensley2015a,Frensley2015c,JunHuang2015}.  The mathematical properties of 
these first-order difference operators will be reviewed below.

Graphene and materials with similar energy-band structures present the problem of the spurious dispersion in its
starkest form.  It is desired to treat these materials with a simple massless Dirac Hamiltonian:\cite{CastroNeto2009}
\begin{equation} \label{eqn:masslessDirac}
 \hat{H} = \hbar v_F \hat{\bf K} \cdot{\bf \sigma},
\end{equation}
where $\hat{\bf K} = -i\nabla$ and ${\bf \sigma}$ is the vector of Pauli spin matrices.  It should not be surprising that
the pitfalls of discretizing this equation have been encountered in the study of quantum field theory on discrete domains, 
where it is known as the ``fermion doubling problem,'' as pointed out by Masum Habib, Sajjad and Ghosh\cite{MasumHabib2015}.
The key insight from the field-theory literature is provided by Stacy\cite{Stacy1982}.  He considered the first-order difference
models and concluded that they could work in one dimension, but in three dimensions the $\hat{K}_z \sigma_z$ term could not be
made hermitian.  Stacy did not explicitly consider the two-dimensional case, but the implication is clear that it can succeed.
The two-dimensional theory is developed in detail in the present work.

The aproach employed in this paper differs from convention in that it treats the spatially discrete formulation as a fully valid
physical theory, not as an approximation to a continuum theory.  Traditionally, continuum theories have been formulated with 
due regard for self-consistency.  Discrete models are viewed as approximations valid only to some asymptotic order and, because such models
suffer from ``error,'' inconsistencies of higher order are only to be expected.  Such notions have created a rich legacy of 
numerical difficulties: instabilities, spurious states, and the need to adjust parameters to avoid those problems.  All of these
difficulties are entirely avoidable.  The present work is offered as an example of how a discrete theory, adapted to numerical
computation, may be constructed by taking self-consistency as the overriding principle.  
As a consequence, the presentation is a bit more tutorial than is often expected.  It also 
addresses aspects of the model that are essential for realistic numerical computations, in particular the open boundary conditions.

\section{Fundamentals of the Discrete Formulation}

  We begin by considering models defined on a one-dimensional discrete space with points uniformly
separated by a distance $a$, so that the position of a point $x_n = na$.  

We define three different representations of the gradient $\hat{K} = -i\partial_x$:
\begin{subequations}
\begin{align}
 \left(\hat{K}_C f \right)_n &= \frac{-i}{2a} \left( f_{n+1} - f_{n-1} \right) , \label{eqn:Kcentered} \\
 \left(\hat{K}_L f \right)_n &= \frac{-i}{a} \left( f_{n} - f_{n-1} \right) , \\
 \left(\hat{K}_R f \right)_n &= \frac{-i}{a} \left( f_{n+1} - f_{n} \right) .
\end{align}
\end{subequations}
Also important is the (far more unique) discretization of the Laplacian $-\partial_x^2$:
\begin{equation}  \label{eqn:Laplacian}
 \left(\hat{L} f \right)_n = \frac{1}{a^2} \left( -f_{n-1} + 2 f_n - f_{n+1} \right). 
\end{equation}

These operators obey some simple relations.  The first-order differences, $\hat{K}_L$ and $\hat{K}_R$
form an adjoint pair:
\begin{equation}  \label{eqn:KLRadjoints}
 \hat{K}_L^\dagger = \hat{K}_R.
\end{equation}
Also,
\begin{equation}
  \hat{K}_L \hat{K}_R = \hat{K}_R \hat{K}_L = \hat{L},  \label{eqn:Ldecomp}
\end{equation}
exactly for an unbounded system.  This may be readily verified by direct calculation.  On the other hand,
\begin{equation}
 \hat{K}_C^2 \neq \hat{L}.
\end{equation}

\subsection{One-Dimensional Model for a Graphene-Like Dispersion Relation}

\begin{figure}
 \centering
 \includegraphics[width=2.3in]{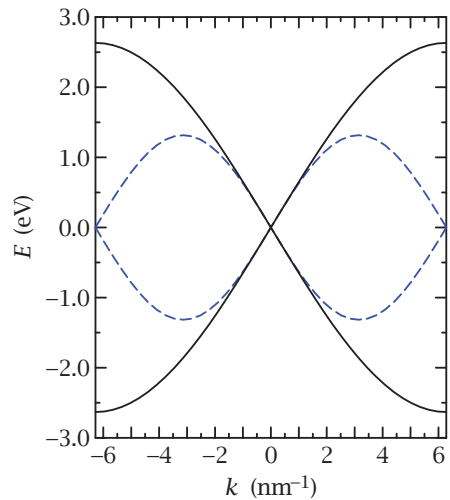}
 \caption{Discrete band structures for the simple 2-band ${\bf k}\cdot{\bf p}$ model using first and second order discretizations.  
 The first-order discretization produces the solid line-curve, and the centered-difference produces the dashed-line curve. The mesh
 spacing is 0.5 nm.}
 \label{fig:kp2band}
\end{figure}

The $x$-dependent term of the Dirac Hamiltonian (\ref{eqn:masslessDirac}) is:
\begin{equation}
 \hat{H}_x = \hbar\hbar v_F \hat{K}_x \sigma_x = \hbar v_F \begin{bmatrix} 0 &   k \\   k & 0 \end{bmatrix},
\end{equation} 
which is also the zero-gap case of the two-band ${\bf k}\mathord{\cdot}{\bf p}$ model considered in Ref.\ \onlinecite{Frensley2015a}.
The continuum approximation produces a dispersion relation:
\begin{equation}
 E(k) = \pm \hbar v_F k.
\end{equation}
The Hamiltonian may be discretized using a second-order formulation:
\begin{gather}
  \Hat{H}_{xC} =  \hbar v_F \begin{bmatrix} 0 & \hat{K}_C   \\   \hat{K}_C & 0 \end{bmatrix}  \label{eqn:HxC}
\intertext{ or using a first-order difference: }
  \hat{H}_{xL} =  \hbar v_F \begin{bmatrix}  0 & \hat{K}_R  \\  \hat{K}_L & 0 \end{bmatrix} .  \label{eqn:HxL}
\end{gather}
Applying these forms to a plane wave yields the bandstructure:
\begin{equation} \label{eqn:gen2bandEk}
 E(k) = \pm \frac{\hbar v_F}{a} \sin(ka), 
\end{equation}
for the centered difference, but
\begin{align}
 E(k) &= \pm \frac{\hbar v_F}{a} \sqrt{2 \left[1 - \cos(ka) \right] } \notag \\
 &= \pm \frac{2\hbar v_F}{a} \sin(ka/2),  \label{eqn:dispRel1}
\end{align}
from the first-order Hamiltonian.
The resulting functions are plotted in
Fig.\ \ref{fig:kp2band}.

\subsection{Eigenstates of a Bounded System}

When we examine the wavefunctions produced by the discrete models, we first need to address the point that (\ref{eqn:HxL})
is not unique.  We could equally well have written:
\begin{equation}
  \hat{H}_{xR} =  \hbar v_F \begin{bmatrix}  0 & \hat{K}_L  \\  \hat{K}_R & 0 \end{bmatrix} .  \label{eqn:HxR}
\end{equation}
The subscripts L,R refer to the operator placed in the lower triangle of the pseudospin-based matrix.  
We will write the pseudospin wavefunction at a point $j$ as:
\begin{equation*}
 \Psi_j = \begin{bmatrix} u_j \\ v_j \end{bmatrix} ,
\end{equation*}
and, for purposes of numerical efficiency, assume a mapping from the paired spatial and pseudospin indices to a single index
in which the psuedospin index varies most rapidly.  This minimizes the matrix bandwidth, as opposed to ordering schemes in which
the spatial index is taken to be the rapidly varying one\cite{Jiang2014}.

\begin{figure}
 \centering
  \includegraphics[width=2.5in]{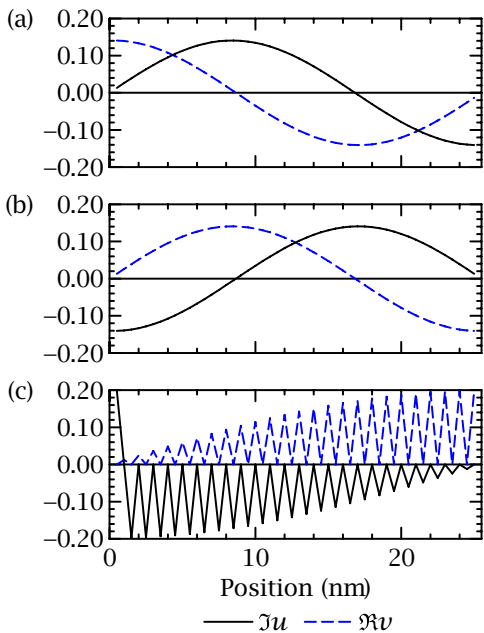}
  \caption{The second state above the Dirac point in a graphene structure of 25 nm width, with $a$ = 0.5 nm.  
 The discrete formulations are:
 (a) left-hand, (b) right-hand and (c) centered difference. Case (c) is clearly spurious.  }
 \label{fig:XboundStates}
\end{figure}

The eigenstates of a typical energy level near the Dirac point are illustrated in Fig.\ \ref{fig:XboundStates}, for a system
in which the Hamiltonian has simply been truncated at the boundaries, for each of the Hamiltonians (\ref{eqn:HxC}),
(\ref{eqn:HxL}) and (\ref{eqn:HxR}).
While the first-order Hamiltonians produce plausible wavefunctions, (a) and (b), the centered-difference (c) shows the
expected contamination by spurious short-wavelength components.  The energies obtained from the first-order Hamiltonians 
coincide to within the expected numerical precision, while those obtained from the centered-difference approximation differ quite markedly.

The wavefunctions in Fig.\ \ref{fig:XboundStates} (a) and (b) show a clear symmetry relation, and an apparent quadrature phase
relation between the pseudospin components.  This holds for states near the Dirac point; at more extreme energies the phase
relations approach closer to 0 or $\pi$.  Also note that the probability density does not approach zero at either boundary.
We cannot impose Dirichlet boundary conditions on both boundaries of a first-order operator.

The Hamiltonians $\hat{H}_{xL}$ and $\hat{H}_{xR}$ are in fact related by a similarity transformation:
\begin{equation}  \label{eqn:sigmaSymmetry}
  \hat{H}_{xR} = \sigma_x \hat{H}_{xL} \sigma_x,
\end{equation}
where the identity operator with respect to the spatial indices is implied in the Pauli matrix.  This may be
verified by direct calculation.  

\subsection{The Second Dimension}

\begin{figure}
 \centering
  \includegraphics[width=2.5in]{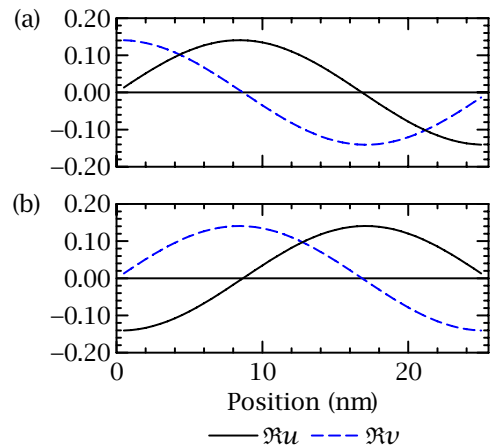}
  \caption{Eigenstate solutions of $H_y$ for the same conditions as in Fig.\ \ref{fig:XboundStates}.  
 Again, the discrete formulations are:
 (a) left-hand and (b) right-hand.  }
 \label{fig:YboundStates}
\end{figure}

Hamiltonians for the y-dependent term of (\ref{eqn:masslessDirac}) may be analogously defined:
\begin{gather}
  \hat{H}_{yL} =  \hbar v_F \begin{bmatrix}  0 & -i \hat{K}_R  \\  i \hat{K}_L & 0 \end{bmatrix} ,  \label{eqn:HyL}
\intertext{and}
  \hat{H}_{yR} =  \hbar v_F \begin{bmatrix}  0 & -i \hat{K}_L  \\  i\hat{K}_R & 0 \end{bmatrix} .  \label{eqn:HyR}
\end{gather}
The elements of these operators are purely real, and therefore their eigenstates will also be real-valued.
The eigenstates corresponding to the simulation of Fig.\ \ref{fig:XboundStates} are shown in Fig.\ \ref{fig:YboundStates}.
Also,
\begin{equation} \label{eqn:sigmaYsymmetry}
  \hat{H}_{yR} = \sigma_y \hat{H}_{yL} \sigma_y,
\end{equation}
as one would expect from (\ref{eqn:sigmaSymmetry}).

\subsection{Time Reversal}

The time reversal operation $\cal{T}$ which leaves a Dirac equaiton of the form (\ref{eqn:masslessDirac}) invariant consists of 
complex conjugation concatenated with the unitary transformation produced by $\sigma_y$.
When this operation is applied to our Hamiltonians, we find that the left and right formulations are simply interchanged:
\begin{subequations}\begin{align}
{\cal T}(\hat{H}_{xL} ) &= \hat{H}_{xR}, \\
{\cal T}(\hat{H}_{yL} ) &= \hat{H}_{yR}.
\end{align}\end{subequations}
Because the $\hat{H}_{L}$ and $\hat{H}_{R}$ are also related by the unitary transformations (\ref{eqn:sigmaSymmetry}) and (\ref{eqn:sigmaYsymmetry}),
the left and right formulations must produce identical results for any physical observable (for an unbounded system).  Thus, a 
time-reversed formulation will also produce identical results.

\subsection{Propagating States}

\begin{figure}
 \centering
  \includegraphics[width=2.5in]{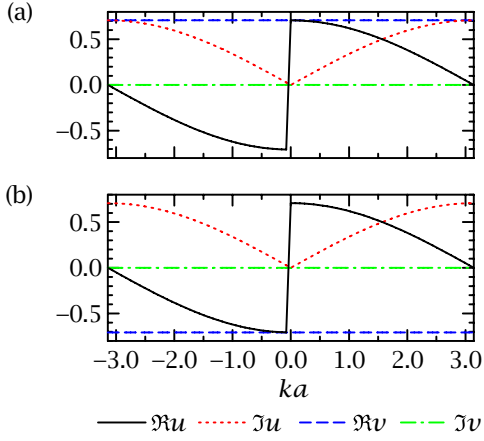}
  \caption{Coefficients of the plane-wave states from $\hat{H}_{xL}$ as functions of $k$. 
 (a) $E>0$ (b) $E<0$  }
 \label{fig:coefXL}
\end{figure}

\begin{figure}
 \centering
  \includegraphics[width=2.5in]{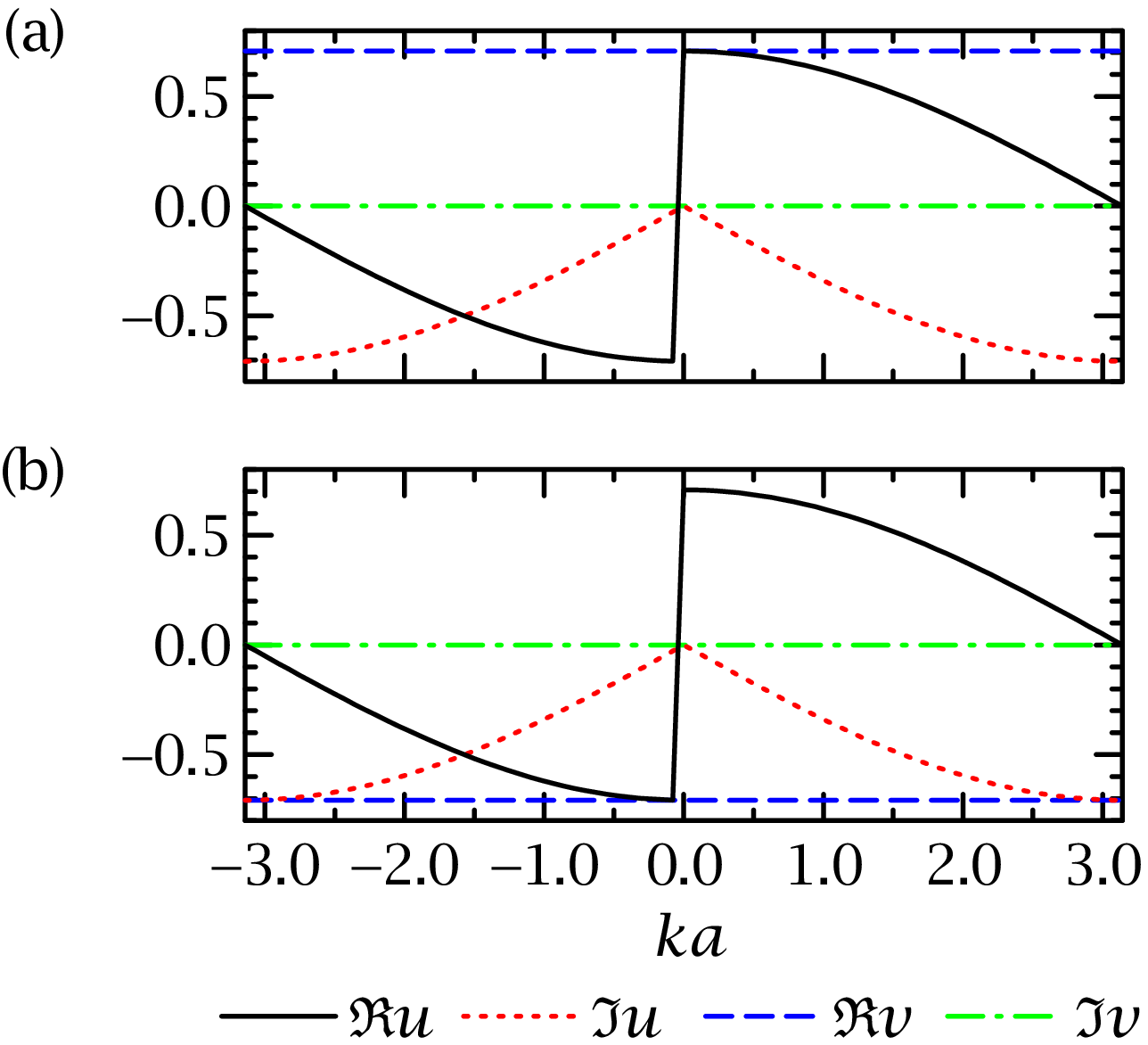}
  \caption{Coefficients of the plane-wave states from $\hat{H}_{xR}$ as functions of $k$. 
 (a) $E>0$ (b) $E<0$  }
 \label{fig:coefXR}
\end{figure}

\begin{figure}
 \centering
  \includegraphics[width=2.5in]{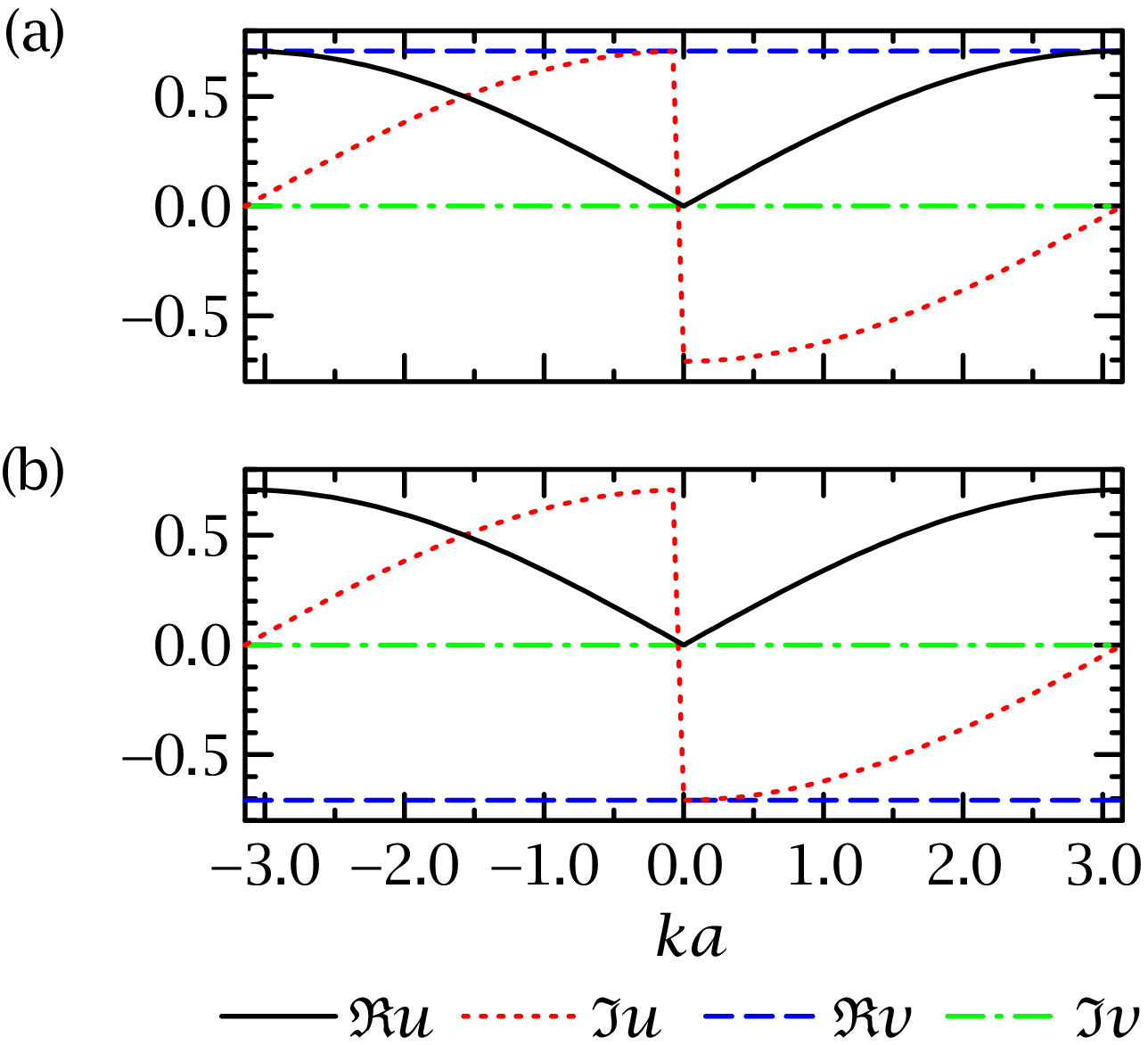}
  \caption{Coefficients of the plane-wave states from $\hat{H}_{yL}$ as functions of $k$. 
 (a) $E>0$ (b) $E<0$  }
 \label{fig:coefYL}
\end{figure}

\begin{figure}
 \centering
  \includegraphics[width=2.5in]{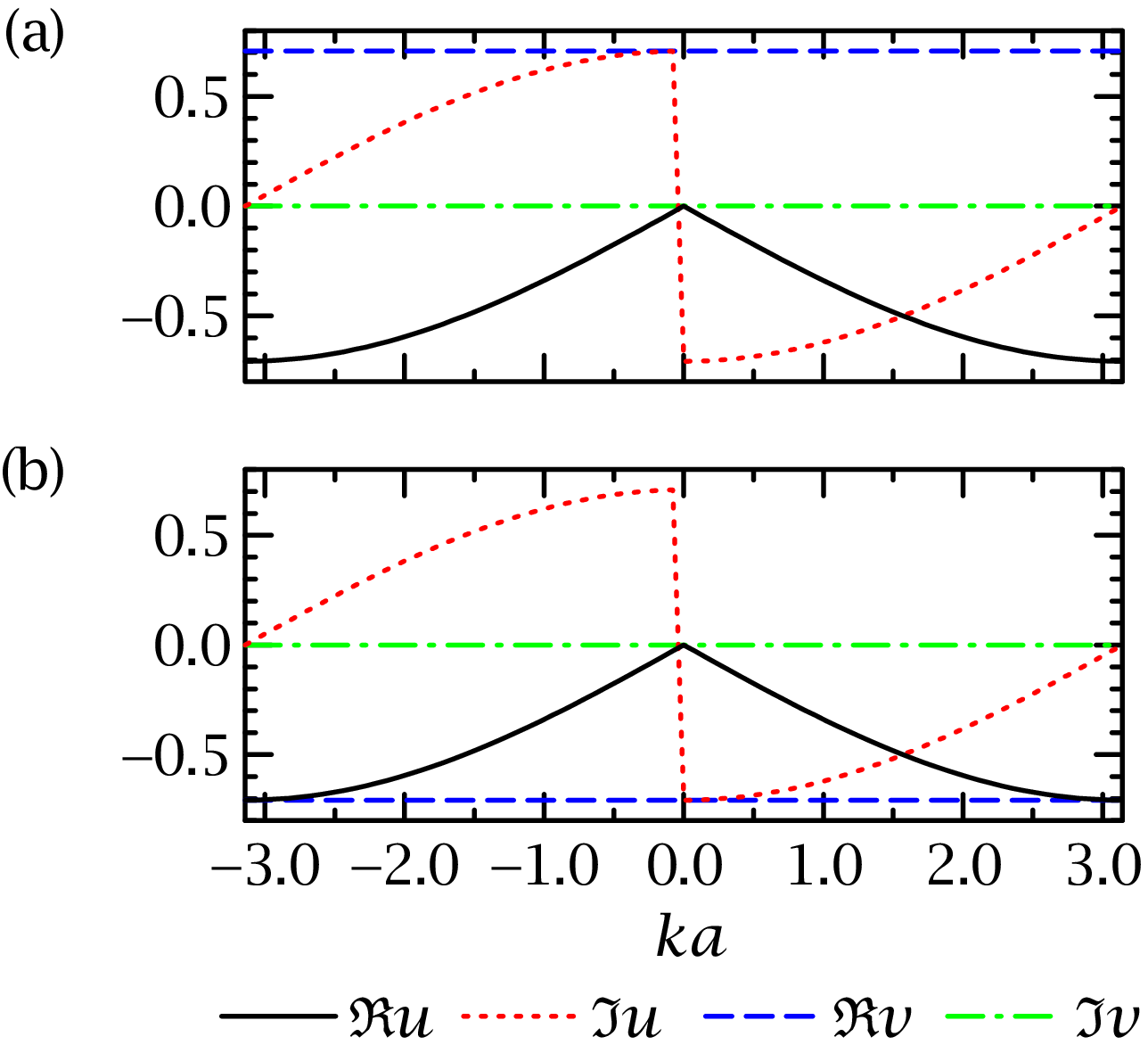}
  \caption{Coefficients of the plane-wave states from $\hat{H}_{yR}$ as functions of $k$. 
 (a) $E>0$ (b) $E<0$  }
 \label{fig:coefYR}
\end{figure}

The propagating states are of the form
\begin{equation}  \label{eqn:planeWavePsi}
 \Psi_j(k) = e^{ikja}  \begin{bmatrix} u(k) \\ v(k) \end{bmatrix},
\end{equation}
and there will be two independent solutions at each $k$, with positive and negative energies.
The most straightforward way to study these solutions is to simply insert (\ref{eqn:planeWavePsi}) into each of our
Hamiltonians and then numerically solve the resulting eigensystem to find $u(k)$ and $v(k)$.  There is, of course, 
an overall phase ambiguity, but we will find the relative phase relations, and verify that the magnitudes are independent of 
$k$.  The results of such a procedure are shown in Figs.\ \ref{fig:coefXL}--\ref{fig:coefYR}.

The propagating plane-wave states for $\hat{H}_{xL}$ and $\hat{H}_{xR}$ can be expressed as:
\begin{equation}
  \Psi_j = \frac{e^{ikja}}{\sqrt{2}} \begin{bmatrix} \sgn(k)\cos(ka/2) \pm i \left| \sin(ka/2)\right|   \\  
     \sgn(E) \times 1   \end{bmatrix}  .
\end{equation} 
In the $y$-direction, we find:
\begin{equation}
  \Psi_j = \frac{e^{ikja}}{\sqrt{2}} \begin{bmatrix} \pm \left| \sin(ka/2)\right| - i \sgn(k)\cos(ka/2)     \\  
     \sgn(E) \times 1   \end{bmatrix}  ,
\end{equation}
where, in both cases, the $+$ sign applies to eigenstates of $\hat{H}_{yL}$ and the $-$ sign to states of $\hat{H}_{yR}$.

One of the interesting features of Figs.\ \ref{fig:coefXL}--\ref{fig:coefYR} is the abrupt phase reversal of the $u$
components as $k$ passes through 0.  As mentioned, there is an overall phase ambiguity, and if we were to impose a phase change
of $\pi$ at $k=0$ the $u$ functions would then be continuous and there would be a discontinuity in $v$.  But, we could
remove this discontinuity by interchanging the identities of the positive and negative $E$ solutions as $k$ passes through zero.
Thus we see that there is a symmetry that has been obscured by the idiosyncrasies of the numerical eigensystem algorithm.
\footnote{The implicit QL algorithm that is at the heart of all direct solutions routines for Hermitian or symmetric matrices
works from the highest index to the lowest.  Thus the rotations that produce the eigenvectors, and which are accumulated through
the procedure, tend to be simplest for high-index eigenvectors.  But, this can be obscured by permutation of the eigenvectors
as the eigenvalues are sorted in increasing magnitude. } But, the possibility of discontinuities in the plane-wave coefficients
illustrates an important point that will appear again in the Green's function for the system.

\section{Required Tools for Electronic Device Analysis}

For the present purposes an ``electronic device'' is any experimental sample to which electrical contacts have been made, permitting
current measurements.  To simulate such structures one requires an explicit expression for the (preferably local) current density,
a formulation for (preferably reflectionless) open boundary conditions and, for the more interesting devices, a well-defined
treatment of compositionally heterogeneous structures.   

\subsection{Current Density}

The current density expressions implied by these Hamiltonians are properly derived from the discrete Green's identity, rather
than from a velocity operator\cite{Frensley2015b}.  The resulting expressions for the current flowing between adjacent mesh
points $j$ and $j+1$ are:
\begin{subequations}\begin{align}
 \langle J_{xL} \rangle_{j+\text{\textonehalf}} &= \frac{v_F}{a} \left( u_{j}^* v_{j+1} + u_{j} v_{j+1}^* \right), \\
 \langle J_{xR} \rangle_{j+\text{\textonehalf}} &= \frac{v_F}{a} \left( u_{j+1}^* v_{j} + u_{j+1} v_{j}^* \right),  \\
 \langle J_{yL} \rangle_{j+\text{\textonehalf}} &= \frac{i v_F}{a} \left( u_{j} v_{j+1}^* - u_{j}^* v_{j+1} \right), \\
 \langle J_{yR} \rangle_{j+\text{\textonehalf}} &= \frac{i v_F}{a} \left( u_{j+1}^* v_{j} - u_{j+1} v_{j}^* \right) .
\end{align}\end{subequations}

\subsection{Open Boundary Conditions}

The development of open boundary conditions proceeds in the manner described by Datta\cite{Datta1995}.  
For the one-dimensional case, we
use the recursion relation for the diagonal block of the Green's function for a linear chain:
\begin{equation} \label{eqn:Grecursion}
  \hat{G}_{n+1,n+1} = \left[ \hat{D}_{n,n} - \hat{S}_{n+1,n}\hat{G}_{n,n}\hat{S}_{n,n+1} \right]^{-1},
\end{equation}
where $\hat{D}$ is a diagonal block and $\hat{S}$ is an off-diagonal block of the block-structured Schroedinger
operator $E-\hat{H}$.  Equation (\ref{eqn:Grecursion}) is used twice in the process of finding the self-energy $\Sigma$.
First, we solve for the boundary diagonal block of a semi-infinite system by invoking the principle that adding one more
element to an infinite chain of identical elements produces the same semi-infinite system.  Thus, we solve:
\begin{equation}
  \hat{G}_{n+1,n+1} = \hat{G}_{n,n} \equiv \hat{g}.
\end{equation}
To simplify the resulting expressions let
\begin{equation}
  A = \frac{\hbar v_f}{a} .
\end{equation}
Then the element of $\hat{g}$ which turns out to be the one that is needed is:
\begin{equation} \label{eqn:Gbdy}
  g_{uu}(E) = \frac{1}{A} \left[ \frac{E}{2A} \pm \sqrt{\left(\frac{E}{2A}\right)^2 - 1 } \right].
\end{equation}
Observe that this expression has the required branch cut for $-2A \le E \le 2A$, as expected for a continuous energy band.
\footnote{  When numerically evaluating the self-energy using (\ref{eqn:Gbdy}) one needs to be sure that the computer code
does indeed put the branch cut in the proper place.  The usual convention for the complex square root function, placing the
branch cut along the negative real axis, should insure a proper implementation of (\ref{eqn:Gbdy}).  However, the numerical
evaluation is not generally done by conversion to polar form, but by solution of quadratic equations in the cartesian form, and
the branch cut is implicitly placed by conditional expressions which resolve the inherent sign ambiguities.  Thus, the placement
of the branch cut should be explicitly checked.  }

The second use of (\ref{eqn:Grecursion}) is to apply it to the corner diagonal blocks of the system domain, defining the boundary 
self-energy $\Sigma$ as the final term in (\ref{eqn:Grecursion}).  For the present first-order models, there  is only one coupling term
out of the domain at each boundary, and that is the reason only one $\Sigma$ is required.  It is given by
\begin{equation}  \label{eqn:bdySelfEnergy}
  \Sigma(E) = A \left[ \frac{E}{2A} - \sqrt{\left(\frac{E+i\epsilon}{2A}\right)^2 - 1 } \right],
\end{equation} 
where the small positive imaginary part added to $E$ is indicated at the point in the formula where it is needed to obtain the 
retarded Green's function.  
The two $\Sigma$ terms appear in the diagonal elements of the state that is coupled out of the domain.  For example, 
using $\hat{H}_{xL}$ as shown
in (\ref{app:HxL}), 
and if the boundaries are at $j = 1$ and $j = n$, the boundary diagonal blocks become:
\begin{subequations}\begin{gather}
 \left[\hat{H}_{xL}\right]_{11} = 
  \begin{bmatrix} 0 & iA \\ -iA & \Sigma \end{bmatrix} \\
 \left[\hat{H}_{xL}\right]_{nn} = 
  \begin{bmatrix} \Sigma & iA \\ -iA & 0 \end{bmatrix} .
\end{gather}\end{subequations}  

This open-system formulation can be easily tested by explicitly evaluating the retarded Green's function for a finite 
system.  The proper formulation of the self-energy is verified by a solution which shows no evidence of standing waves
due to reflection at the boundaries, and experience has shown that the spurious reflections due to algebraic errors are usually
quite apparent.  We will solve for one particular column of the retarded Green's function, represented
as a wavefunction $\Psi$, by solving:
\begin{equation}  \label{eqn:impulseResponse}
 \left(E - \hat{H} \right) \Psi = \delta_{j,j_0} \delta_{s,s_0},
\end{equation}
where $\hat{H}$ is the effective Hamiltonian including the boundary self-energies, and $s$ represents the pseudospin index.
The corresponding continuum equation would contain a first-order differential operator and a Dirac delta function.  Such
a formulation implies that there must be a discontinuity in the value of the wavefunction, not just a slope discontinuity
as is the case in a conventional second-order Hamiltonian for a massive particle.  The retarded Green's function for a
uniform system should just consist of outgoing waves propagating away from the source of the impulse.  That is, a positive
$k$ plane wave for $j > j_0$ and a negative $k$ plane wave for $j < j_0$.  Thus, we see that the discontinuity in the plane-wave
coefficients seen in Figs.\ \ref{fig:coefXL}--\ref{fig:coefYR} can be reflected in discontinuities in the Green's function.

\begin{figure}
 \centering
  \includegraphics[width=2.5in]{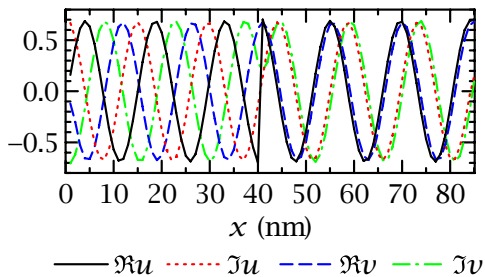}
  \caption{The retarded Green's function for an open uniform system using $\hat{H}_{xL}$.  The energy is 0.1 eV, $a$ = 1.0 nm and the impulse
   is applied at $x = 40$ nm. The purpose of the calculation is to demonstrate that the boundary self-energies have been 
   correctly formulated, as evidenced by the absence of reflections at the domain boundaries.  }
 \label{fig:RetGrnFn}
\end{figure}

A wavefunction computed as described above is shown in Fig.\ \ref{fig:RetGrnFn}.
This does indeed verify the correctness of the open boundary conditions.  There is no evidence of a standing-wave envelope 
which would result from spurious reflections at the boundaries, and there is no apparent restriction on the value of any
component at the boundaries, as compared to the confined states of Figs.\ \ref{fig:XboundStates} and \ref{fig:YboundStates}.  
The wavefunction discontinuity is apparent in the real part of $u$.  This was achieved by a slight manipulation of
(\ref{eqn:impulseResponse}): the impulse was taken to be $i/a$ and applied to the $v$ pseudospin component, to make 
the discontinuity visually apparent.

For a two-dimensional calculation, the self-energy of (\ref{eqn:bdySelfEnergy}) must be applied to each transverse wavefunction
mode $m$.  Then $\Sigma$ becomes a matrix in the transverse coordinates, defined by:
\begin{equation}
 \hat{\Sigma}(E) = \sum_m \Sigma(E-E_m) \hat{P}_m,
\end{equation}
where $E_m$ is the transverse energy associated with mode $m$ and $\hat{P}_m = |m\rangle\langle m |$ is the corresponding 
projection operator.

\subsection{Massive Dirac Particles}

Many of the more recently studied graphene-like systems display a nonzero energy-band gap $E_G$.  These are conventionally included 
by adding a term to the Hamiltonian of the form:
\begin{equation}
 \hat{H}_z = M \sigma_z,
\end{equation}
where $M = \text{\textonehalf} E_G$.
Adding this term to any one of our Hamiltonians $\hat{H}_{x,y}$ produces a model which is just the two-band ${\bf k}\mathord{\cdot}{\bf p}$
model studied in Ref.\ \onlinecite{Frensley2015a}.  The resulting dispersion relation is:
\begin{equation} \label{eqn:massDiscrDispersion}
  E = \pm \sqrt{ M^2 + 4 A^2 \sin^2(ka/2) }  .
\end{equation}
Evaluating the $k=0$ effective mass using
\begin{equation*}
\frac{1}{m^*} = \frac{1}{\hbar^2} \frac{\partial^2 E}{\partial k^2},
\end{equation*}
we find:
\begin{equation}
  M = m^* v_F^2.
\end{equation}

  The dispersion relations (\ref{eqn:massDiscrDispersion}) are monotonic, and their form is illustrated in Fig.\ 2 of 
Ref.\ \onlinecite{Frensley2015a}. The energy bands extend from $E = \pm M$ to $E = \pm \sqrt{M^2+4A^2}$.  Thus, the 
branch cut in the Green's function for an electrical lead [corresponding to (\ref{eqn:Gbdy})] will be split into two
branch cuts, spanning these intervals.

\subsection{Heterostructures}

The author's recent paper\cite{Frensley2015c} has described how to construct consistent discretizations for heterostructures by applying 
the variational formulation of the wave equation to a discretely defined wavefunction.  The problem that was solved is that
the usual procedures of mathematical approximation generally lead to inconsistent treatment of terms of different powers of $k$
in the ${\bf k}\mathord{\cdot}{\bf p}$ Hamiltonian.  These inconsistencies produce various anomalies in the computed wavefunctions,
and the solution to such problems has been sought in terms of the ordering of the operator forms of position-dependent quantities.\cite{Vogl2011}
The problem is really one of mapping the physical structure one wishes to simulate onto the discrete domain of the computation,
and as such needs to be addressed by spatially resolving both the physical heterostructure and the sampled wavefunction.
Also, to repeat the warning of Ref.\ \onlinecite{Frensley2015c}: The first-order differences must be explicitly built into the 
Lagrangian; any formulation that appears to be more ``natural'' will produce the centered-difference form (\ref{eqn:HxC}) with its inherent
instabilities.  

The present problem is sufficiently simple that the results of the variational approach may be explicitly specified.  The essential
point is to recognize that there are two possible conventions for mapping a heterostructure onto a discrete model.  They are
to (a) assume that the heterojunctions are located at the midpoint of a mesh interval, or (b) to assume that the heterojunctions
coincide with a mesh point.  For the graphene-like band structure, the three material parameters that can potentially
change across a heterojunction are the Fermi velocity as contained in the constant $A$, the mass factor $M$ and the energy of the 
Dirac point $E_D$ that acts as a scalar potential and could abruptly shift across such a junction.   

For case (a), The material parameters are associated with the meshpoint, and are taken to be constant across the interval from
$(j-\text{\textonehalf})a$ to $(j+\text{\textonehalf})a$. We denote these values $A_{j}$, $M_j$ and $E_{D;j}$.  Now, because the 
derivative is associated with the interval between adjacent meshpoints, the $\hbar v_F \hat{K}$ terms in the Hamiltonian must be
associated with the appropriate interval, which we label with half-integer indices.  Because each interval potentially contains
equal lengths of material with two different Fermi velocities, we must average that parameter in the Hamiltonian terms.  
These considerations produce the following formulas for the blocks of the Hamiltonian, if we use $\hat{H}_{xL}$:
\begin{subequations}\label{eqn:heteroJa}\begin{gather}
\hat{H}_{xL;j,j-1} = 
 \begin{bmatrix} 0 & 0 \\ \frac{i}{2}\left( A_{j-1} + A_{j} \right) & 0 \end{bmatrix}, \\
\hat{H}_{xL;j,j} =  
 \begin{bmatrix} E_{D;j} + M_j & \frac{i}{2}\left( A_{j} + A_{j+1} \right) 
  \\ \frac{-i}{2}\left( A_{j-1} + A_{j} \right) & E_{D;j} -M_j \end{bmatrix} , \\
\hat{H}_{xL;j,j+1} = 
 \begin{bmatrix} 0 & \frac{-i}{2}\left( A_{j} + A_{j+1} \right) \\ 0 & 0 \end{bmatrix}
\end{gather}\end{subequations}

For case (b) the considerations are reversed.  The heterojunctions occur at the mesh points, and thus the materials
are uniform across the mesh intervals.  We label the material-dependent parameters with half-integer indices, and we 
find that we need to average the diagonal terms $E_D$ and $M$ between the adjacent intervals.  If we define:
\begin{subequations}\label{eqn:heteroJb}
\begin{gather}
\bar{E}_j = \frac{1}{2}\left(E_{D;j-\text{\textonehalf}} + E_{D;j+\text{\textonehalf}} \right),  \\
\bar{M}_j = \frac{1}{2}\left(M_{j-\text{\textonehalf}} + M_{j+\text{\textonehalf}} \right). 
\intertext{Then the Hamiltonian becomes:}
\hat{H}_{xL;j,j-1} = 
 \begin{bmatrix} 0 & 0 \\ \frac{i}{2} A_{j-\text{\textonehalf}} & 0 \end{bmatrix}, \\
\hat{H}_{xL;j,j} =  
 \begin{bmatrix} \bar{E}_j + \bar{M}_j & \frac{i}{2}A_{j+\text{\textonehalf}} 
  \\ \frac{-i}{2}A_{j-\text{\textonehalf}} & \bar{E}_j - \bar{M}_j \end{bmatrix} , \\
\hat{H}_{xL;j,j+1} = 
 \begin{bmatrix} 0 & \frac{-i}{2} A_{j+\text{\textonehalf}} \\ 0 & 0 \end{bmatrix} .
\end{gather}\end{subequations}

\section{Discussion}

As indicated in the Introduction, the development of the discrete theory has not closely followed the continuum formulation.
A primary reason is that the discrete theory contains a much richer mathematical structure than does the continuum version.
This is largely due to the fact that in the discrete theory the energy bands are bounded, and the transition of the
wavefunctions to standing waves near the Brillouin zone boundary must be comprehended.  In the continuum theory
the bands are unbounded, and the form of the wavefunctions is independent of $k$. 

Now, the appropriate question is: Which formulation is the better representation of the physical system?  The term ``accuracy''
is being deliberately avoided, because it has come to mean the discrepancy between continuum and discrete formulations, with
the continuum answer being considered the standard for comparison.  If one considers the energy-band structure of graphene\cite{Wallace1947}
somewhat beyond the immediate neighborhood of the Dirac points, we see dispersion relations that bend over and reach extrema, as
does the discrete formulation.  Envelope function models which are expressed as low-order differential operators 
inevitably produce unbounded band structures, and prompt concerns over the proper mathematical treatment of unbounded operators\cite{Vogl2011}.
No condensed-matter system shows unbounded bands, and no spatially discrete model can ever contain unbounded operators.
It would thus appear to be logical to accept the results of discrete models more readily than the results of continuum models. 

The key to understanding the success of the first-order differencing is to realize that the relation which must
be preserved exactly, in both the continuum and discrete domains is
\begin{equation}  \label{eqn:Ksquared}
 \hat{L} = \hat{K}^2.
\end{equation}
Because the dispersion relation of $\hat{L}$:
\begin{equation}
 L(k) = \frac{2}{a^2} \left[ 1 - \cos(ka) \right],
\end{equation} 
is monotonic within the Brillouin zone, an adjoint pair of operators which act as an exact square root of $\hat{L}$ will
also produce a monotonic dispersion relation, and that is precisely what we found (\ref{eqn:dispRel1}).  Now, it is
impossible to satisfy (\ref{eqn:Ksquared}) in the discrete domain with a single self-adjoint $\hat{K}$.  But, as we have
seen, it is readily satisfied with an adjoint pair of operators.  This situation is reminiscent of those
addressed by Dirac in his analyses of the harmonic oscillator and the relativistic electron equation,
but this problem is vastly simpler, because $\hat{K}_L$ and $\hat{K}_R$ commute.  

In the spirit of Dirac's approach, let us note that 
\begin{equation}
 \left( \hat{H}_x + \hat{H}_y \right)^2 = \hbar^2 v_F^2 \left( \hat{L}_x^2 + \hat{L}_y^2 \right),
\end{equation}
for any choice of left or right handed formulations for $\hat{H}_x$ and $\hat{H}_y$.  This is just the discrete version of the condition
that Dirac imposed upon the relativistic wave equation (neglecting the mass term).

In summary, an unconditionally stable and robust discrete theoretical model of two-dimensional massless Dirac particles can be constructed 
by employing a first-order difference operator in the role of the gradient.  The use of first-order operators entails ambiguities which
appear to violate spatial symmetry.  In fact the choice of the ``left-hand'' or ``right-hand'' operator has no physically 
observable consequences.

\bibliography{FrensleyGrapheneEnvFn2015}

\appendix

\section{Explicit Expressions for the Hamiltonians}

The discrete Hamiltonians in each of the $x$ and $y$ directions consist of block-tridiagonal
matrices, where the blocks have a dimension of two, corresponding to the pseudospin degrees of freedom.
The Hamiltonian can thus be specified by writing the three blocks of a general row:   
\begin{equation}  \label{app:HxL}
\hat{H}_{xL} = \frac{\hbar v_F}{a} 
 \begin{bmatrix} 0 & 0 \\ i & 0 \end{bmatrix}
 \begin{bmatrix} 0 & i \\ -i & 0 \end{bmatrix}
 \begin{bmatrix} 0 & -i \\ 0 & 0 \end{bmatrix}.
\end{equation}
\begin{equation}  \label{app:HxR}
\hat{H}_{xR} = \frac{\hbar v_F}{a} 
 \begin{bmatrix} 0 & i \\ 0 & 0 \end{bmatrix}
 \begin{bmatrix} 0 & -i \\ i & 0 \end{bmatrix}
 \begin{bmatrix} 0 & 0 \\ -i & 0 \end{bmatrix}.
\end{equation}
\begin{equation}  \label{app:HxC}
\hat{H}_{xC} = \frac{\hbar v_F}{2a} 
 \begin{bmatrix}0 & i \\ i & 0 \end{bmatrix}
 \begin{bmatrix} 0 & 0 \\ 0 & 0 \end{bmatrix}
 \begin{bmatrix} 0 & -i \\ -i & 0 \end{bmatrix}.
\end{equation}
\begin{equation}  \label{app:HyL}
\hat{H}_{yL} = \frac{\hbar v_F}{a} 
 \begin{bmatrix} 0 & 0 \\ -1 & 0 \end{bmatrix}
 \begin{bmatrix} 0 & 1 \\ 1 & 0 \end{bmatrix}
 \begin{bmatrix} 0 & -1 \\ 0 & 0 \end{bmatrix}.
\end{equation}

\begin{equation}  \label{app:HyR}
\hat{H}_{yR} = \frac{\hbar v_F}{a} 
 \begin{bmatrix} 0 & 1 \\ 0 & 0 \end{bmatrix}
 \begin{bmatrix} 0 & -1 \\ -1 & 0 \end{bmatrix}
 \begin{bmatrix} 0 & 0 \\ 1 & 0 \end{bmatrix}.
\end{equation}

\end{document}